
\magnification 1200  \baselineskip=18pt
 \hfuzz=5pt \null\vskip 2cm
\centerline{{\bf CLASSICAL COULOMB SYSTEMS :}}
\centerline{{\bf SCREENING AND CORRELATIONS REVISITED }}
\bigskip
\centerline {B. Jancovici\footnote{$^1$}{Laboratoire de Physique
Th\'eorique et Hautes Energies,  B\^atiment 211,
Universit\'e de Paris-Sud, 91405 Orsay, France (this
laboratory is  associated with the Centre National de la Recherche
Scientifique).}} \bigskip

\noindent {\bf Abstract}

 From the laws of macroscopic electrostatics of conductors (in
particular the existence of screening) taken for granted, one can
deduce universal properties for the thermal fluctuations in a
classical Coulomb system at equilibrium. The universality is
especially apparent in the long-range correlations of the electrical
potentials and fields. The charge fluctuations are derived from the
field fluctuations. This is a convenient way for studying the surface
charge fluctuations on a conductor with boundaries. Explicit results
are given for simple geometries. The potentials and the fields have
Gaussian fluctuations, except for a short-distance cutoff.


\vfill\eject
\noindent{\bf 1. INTRODUCTION}
\bigskip
A salient property of matter is the screening effect : matter ``in
thermal equilibrium does not tolerate any charge inhomogeneity  over
more than a few intermolecular distances''.$^{(1)}$  In the present
paper, we consider those systems which can be described as made of
charged particles, interacting through Coulomb's law, to which
equilibrium classical (i.e. non-quantum) statistical mechanics is
applicable (for instance, electrolytes); then, screening has
especially rich and simple consequences. From the existence of
screening, taken as granted, it is possible to deduce quantitative
properties of the correlation functions. For instance, in the bulk,
the charge-charge correlation function obeys the well-known
Stillinger-Lovett sum rules;$^{(2)}$ other quantitative properties
hold near walls or interfaces.$^{(3)}$ There is a delicate interplay
between the statistical mechanics of correlations and the validity
of macroscopic electrostatics.

In the standard approaches, the focus is on the charges. One assumes
that an external charge is screened and, using linear response
theory, one obtains information about the charge correlations. The
purpose of the present paper is to revisit the subject with the focus
put on the electrical potential, the electrical field, and their
correlations. From this alternative point of view, it is possible to
rederive known results in a way that we believe to be often simpler
and also to obtain new results. The present method , which relies on
the validity of macroscopic electrostatics, is especially appropriate
for studying surface properties such as surface charge correlations.

The Coulomb systems considered in the present paper form a large
class of models. The one-component plasma is a model such that
identical particles of one sign (say positive) are immersed in a
neutralizing inert continuous background of the opposite sign. The
two-component plasma is made of two species of oppositely charged
particles; a classical theory is possible only if some short-range
interaction is also present, to prevent oppositely charged particles
to collapse on each other. One could also consider more complicated
models, with more than two species of particles, a non-uniform
one-particle potential added, etc..~. We shall only require that the
model be a conductor, in that sense that the laws of macroscopic
electrostatics are assumed to be obeyed for length scales  large
compared to the microscopic characteristic lengths of the model; for
instance, an additional localized charge $Q$ brought into the fluid
should be perfectly screened, i.e. surround itself with a microscopic
polarization cloud of charge $-Q$.

We shall also consider  models with a two-dimensional Coulomb
interaction : in two dimensions, the potential at a distance $r$ from
a unit point charge is $-\ln (r/L)$ (where $L$ is some fixed length)
instead of the familiar $1/r$ which holds in three dimensions.
Two-dimensional Coulomb systems made of oppositely charged particles
exhibit the Kosterlitz-Thouless phase transition :$^{(4)}$ while they
are conductors above some transition temperature, they become
dielectrics below that temperature. A similar transition occurs for
one-dimensional systems with a logarithmic interaction.$^{(5)}$ The
considerations of the present paper apply only to the conducting
phase, i.e. above the transition temperature.

The present approach aims to provide an easy to visualize physically
reasonable picture of classical Coulomb fluids. No attempt is made
towards mathematical rigor. Since a basic ingredient is macroscopic
electrostatics, a number of words will be used with a macroscopic
meaning. ``Inside'' or ``outside'' a Coulomb fluid will mean ``at a
distance from the walls large compared to the microscopic scale''.
``Surface charge'' will mean ``charge in a layer of microscopic
thickness under the surface''. Etc...

The two-point correlations inside and outside a three-dimensional
Coulomb system are discussed in Sections 2 and 3, respectively. In
Section 4, we establish a general method for deriving the surface
charge correlations. Examples are given in Section 5. Section 6 is
about conducting surfaces. Section 7 is about systems with the
two-dimensional logarithmic Coulomb interaction. In Section 8, it is
shown that potential and field fluctuations are Gaussian (except for
microscopic distances).
\bigskip
\noindent{\bf 2. INSIDE A COULOMB SYSTEM}
\bigskip
In this Section, we consider some region inside a three-dimensional
classical Coulomb fluid at equilibrium. This fluid is not necessarily
homogeneous.
\vfill\eject
\noindent {\bf 2.1. Potential and Field Correlations}
\medskip
Let $\phi({\bf r})$ be the microscopic electrical potential at point
${\bf r}$. We wish to rederive the asymptotic formula for the
potential-potential correlation function$^{(6)}$
$$
\beta \langle \phi({\bf r}) \phi ({\bf r'}) \rangle^T \sim {1\over
|{\bf r} - {\bf r'}|} \qquad|{\bf r} -{\bf r'}| \rightarrow \infty
\eqno (2.1) $$
where $\beta$ is the inverse temperature and $\langle \ldots\rangle^T$
means a truncated equilibrium statistical average : $\langle
AB\rangle^T = \langle A B \rangle - \langle A\rangle \langle B \rangle
$.

For deriving (2.1), we use screening and linear response. Let us put
into the fluid an infinitesimal test charge $q$ at ${\bf r}$. The
Hamiltonian $H_0$ of the fluid must now be supplemented with the
fluid-test charge interaction term which can be written as $H' = q
\phi({\bf r})$, where $\phi$ is the potential due to the fluid; it
should be noted that $\phi$ is defined as ${\it not}$ including the
potential due to $q$. By linear response theory, the average of this
potential at some point ${\bf r'}$ is changed by
$$
\langle \phi ({\bf r'}) \rangle_q - \langle \phi ({\bf r'})\rangle =
- \beta q \langle \phi({\bf r}) \phi({\bf r'}) \rangle^T \eqno (2.2)
$$
where   $<\ldots>_q$ means
a statistical average computed with the full Hamiltonian $H_0 + H'$,
while $<\ldots>$ is an average computed with $H_0$ only. Our
assumption about screening is that $q$ surrounds itself with a
polarisation cloud of microscopic size, of charge $-q$; the rest of
the system is unchanged, except that, in the case of an insulated
conductor, charge conservation requires that a charge $+q$ spreads on
the walls. Therefore, for $|{\bf r'} - {\bf r}|$ large compared to
the microscopic scale, the l.h.s. of (2.2) is the potential due to a
point-like charge $-q$ at ${\bf r}$ and a surface charge $+q$; at
${\bf r'}$ inside the conductor, that surface charge creates a
constant potential $q/C$, where $C$ is the capacitance. Altogether,
for an insulated conductor,
$$
\langle \phi ({\bf r'}) \rangle_q - \langle \phi({\bf r'}) \rangle =
- {q\over |{\bf r} - {\bf r'}|} + {q \over C} \eqno (2.3)
$$
{}From (2.2) and (2.3), one obtains
$$
\beta \langle \phi({\bf r}) \phi({\bf r'})\rangle^T = {1\over |{\bf r}
- {\bf r'}|} - {1\over C} \eqno (2.4)
$$
For a grounded conductor, no surface charge appears, and the last
term in (2.3) and (2.4) must be omitted. If the system becomes
infinitely large, $1/C$ goes to zero. Thus, for infinite systems,
one recovers (2.1) in all cases.

 From our derivation, the ``asymptotic'' validity of (2.1) (for an
infinite system) now has a more explicit meaning : (2.1) can be
considered as an equality provided $|{\bf r}- {\bf r'}|$ is large
compared to the microscopic scale (the screening length). From now
on, we shall write (2.1) and similar equations as equalities, with
the understanding that there is some microscopic short-distance
cutoff. This point of view, commonly used in field theory or in the
theory of critical phenomena, is convenient for studying those
properties which are independent of the microscopic detail, i.e.
universal.

Since the electrical field is $E_\mu({\bf r}) = -\partial_\mu
\phi({\bf r})$, the field-field correlation function is easily
obtained from (2.4) (with or without the last term) as
$$
\beta \langle E_\mu ({\bf r}) E_\nu({\bf r'}) \rangle^T =
{\partial^2 \over \partial r_\mu \partial r'_\nu} {1\over |{\bf r}-
{\bf r'}|} = - {3({\bf r}- {\bf r'})_\mu ({\bf r}- {\bf r'})_\nu -
\delta_{\mu \nu}({\bf r}- {\bf r'})^2\over |{\bf r}- {\bf r'}|^5}
\eqno (2.5)
$$
while $\langle {\bf E} ({\bf r}) \rangle = 0$ since $\langle\phi({\bf
r})\rangle$ is a constant  inside a conductor.
\medskip
\noindent{\bf 2.2. Charge Correlations}
\medskip
The charge density $\rho({\bf r})$ is related to the potential
$\phi({\bf r})$ by Poisson's equation $\Delta \phi = - 4 \pi \rho$.
Thus, by taking the Laplacian on ${\bf r}$ in both sides of (2.4),
one obtains
$$
\beta \langle \rho({\bf r}) \phi ({\bf r'}) \rangle^T = \delta({\bf
r}- {\bf r'}) \eqno (2.6)
$$
Of course, writing the r.h.s. of (2.6) as a delta function
disregards some spreading and microscopic structure for small
distances. But the simplified form (2.6) is enough for correctly
giving, by integration upon ${\bf r}$,
$$
\beta \int d{\bf r} \langle \rho ({\bf r}) \phi ({\bf r'}) \rangle^T
= 1 \eqno (2.7)
$$
which is the known Carnie-Chan sum rule.$^{(7)}$

In the case of an homogeneous fluid, the Carnie-Chan sum rule is
equivalent to the Stillinger-Lovett$^{(2)}$ sum rule. We can retrieve
that equivalence by taking the Laplacian on ${\bf r'}$ in both
sides of (2.7), with the result
$$
- 4\pi \beta \langle \rho ({\bf r}) \rho({\bf r'})\rangle^T = \Delta
\delta({\bf r} - {\bf r'}) \eqno (2.8)
$$
(Here the assumption that the fluid is homogeneous is necessary,
because $\delta$ actually stands for some peaked function of ${\bf
r} - {\bf r'}$; if the fluid were non-homogeneous, the microscopic
width of that peak would be a function of ${\bf r'}$, and taking the
Laplacian on ${\bf r'}$ would generate additional terms). From
(2.8), one obtains
$$
\int \langle \rho ({\bf r}) \rho ({\bf r'}) \rangle^T d{\bf r'} = 0
\eqno (2.9a)
$$
and (after an integration by parts)
$$
- {2 \pi \beta\over 3} \int \langle \rho ({\bf r}) \rho ({\bf r'})
\rangle^T ({\bf r}-{\bf r'})^2 d ({\bf r}-{\bf r'}) = 1 \eqno (2.9b)
$$
Eqs (2.9a) and (2.9b) are the well-known Stillinger-Lovett sum rules;
it is amusing that they can be written in the form (2.8), which,
again, disregards the microscopic detail.
\bigskip
\noindent {\bf 3. OUTSIDE A COULOMB SYSTEM}
\bigskip
We now consider correlations involving at least one point outside the
boundaries of a conductor.
\medskip
\noindent {\bf 3.1. Potential and Field Correlations across the
Boundary}
\medskip
We want to compute the potential-potential correlation $\langle \phi
({\bf r}) \phi ({\bf r'}) \rangle^T$ when ${\bf r}$ is inside the
Coulomb fluid and ${\bf r'}$ outside. We can repeat the reasoning of
Section 2.1, considering the response to an infinitesimal test charge
$q$ at ${\bf r}$. The only difference is that the induced surface
charge $q$, which appears in the case of an insulated conductor,
creates at ${\bf r'}$ outside the conductor a potential $q F({\bf
r'})$ rather than $q/C$. This function $F({\bf r'})$ is determined by
macroscopic electrostatics : it is the potential created at ${\bf r'}$
when the conductor carries the total charge $1$. Its explicit form
can be obtained for simple shapes of the conductor.\footnote{$^2$}{In
most textbooks about electrostatics, it is stated that the potential
is continuous at the surface of a conductor. This would imply that
(3.1) and (2.4) become identical when ${\bf
r'}$ is on the surface. Actually, there is usually a potential
difference across the surface of a conductor, due to the formation of
an electrical double layer. However, a constant shift of the
potential is of no effect on the electrical field.} From (2.2), we
now obtain
$$
\beta \langle \phi ({\bf r}) \phi ({\bf r'}) \rangle^T = {1 \over |{\bf
r}- {\bf r'}|} - F(\bf r')\eqno (3.1)
$$
The same result can be obtained by assuming that the test charge is
put at ${\bf r'}$; the total potential of the conductor then is $q
F({\bf r'})$, because $F$ can be regarded as an element of an inverse
capacitance matrix, which is known to be symmetrical.

The last term of (3.1) must be omitted in the case of a grounded
conductor. It vanishes anyhow in the limit of an infinitely large
conductor (such as, for instance, a conducting fluid filling a
half-space).

 From (3.1), one finds a field-field correlation which is again (2.3).
\medskip
\noindent{\bf 3.2. Potential and Field Correlations Outside}
\medskip
Finally, we want to compute the potential-potential corelation
$\langle\phi ({\bf r}) \phi ({\bf r'})\rangle^T$ when both points
${\bf r}$ and ${\bf r'}$ are outside the Coulomb fluid. We still
consider the response to an infinitesimal test charge $q$ at ${\bf
r}$. The total potential change at ${\bf r'}$ is some function $q
G({\bf r}, {\bf r'})$, with $G({\bf r} , {\bf r'})$ determined by
macroscopic electrostatics : $G({\bf r}, {\bf r'})$ is the total
potential change at ${\bf r'}$ when a unit point charge is put at
${\bf r}$. In the case of a finite conductor, for determining $G$ one
must specify whether the conductor is insulated or grounded. $G$ is
explicitly computable for simple shapes of the conductor.

The potential change due to the fluid only is $q [G({\bf r}, {\bf
r'}) - |{\bf r} - {\bf r'}|^{-1}],$ and (2.2) gives
$$
\beta \langle \phi ({\bf r}) \phi({\bf r'}) \rangle^T = {1 \over
|{\bf r} - {\bf r'}|} - G({\bf r}, {\bf r'}) \eqno (3.2)
$$
{}From (3.2), one finds for the field-field correlation
$$
\beta \langle E_\mu ({\bf r}) E_\nu({\bf r'}) \rangle^T = {\partial^2
\over \partial r_\mu \partial r'_\nu} \big [ {1 \over |{\bf r} - {\bf
r'}|} - G ({\bf r}, {\bf r'})] \eqno (3.3)
$$
\bigskip
\noindent{\bf 4. SURFACE CHARGE CORRELATIONS}
\bigskip
In the present macroscopic approach, it is natural to introduce a
surface charge density $\sigma$, which will be associated to the
electrical field discontinuity at the surface of the conductor. At
some point ${\bf r}$ on the surface,
$$
4\pi \sigma ({\bf r}) = E^{\hbox{out}}_n ({\bf r}) -
E^{\hbox{in}}_n({\bf r}) \eqno (4.1.)
$$
where the index $n$ denotes the component normal to the surface (with
the positive direction defined as pointing towards the outside) and
$E_n^{\hbox{out (in)}}({\bf r})$ is the limit of that field component
as ${\bf r}$ is approached from the outside (inside).
Therefore, the surface charge density correlation function is
$$
\langle \sigma ({\bf r}) \sigma({\bf r'}) \rangle^T = {1 \over
(4\pi)^2} \langle [E^{\hbox{out}}_n ({\bf r}) - E^{\hbox{in}}_n
({\bf r})] [E^{\hbox{out}}_n({\bf r'}) - E^{\hbox{in}}_n({\bf
r'})]\rangle^T \eqno (4.2)
$$
Using the field correlations (2.5) and (3.3), one finds the final
result
$$
\beta \langle \sigma ({\bf r}) \sigma ({\bf r'})\rangle^T = - {1\over
(4\pi )^2} {\partial^2 G({\bf r}, {\bf r'})\over \partial
r_n \partial  r'_n} \big |_{{\bf r}, {\bf r'} \in
\hbox{surface}}  \eqno (4.3)
$$

Of course, this approach is valid only if the distance $|{\bf r} -
{\bf r'}|$ is large compared to the microscopic scale. Also, the
surface charge density $\sigma$ has to be understood as being the
microscopic volume charge density, integrated on some microscopic
depth. Within these limitations, the computation of the correlation
function (4.3) has been reduced to a problem in macroscopic
electrostatics of conductors.

A method for obtaining the charge correlations near the surface of a
Coulomb system has been devised by Choquard {\it et al}.$^{(8)}$ a few
years ago. This method was presented as an approximation to the
Debye-H\"uckel approximation. However, it can be seen that the surface
charge densities provided by the method of Choquard {\it et al}. have
to be identical to ours (their kernel $G_{\hbox{ext}}$ is identical to
our function $G$). Therefore, our approach rephrases the method of
Choquard {\it et al}., without any need for invoking the Debye-H\"uckel
approximation.
\bigskip
\noindent {\bf 5. A FEW SPECIFIC CASES}
\bigskip
In order to illustrate how the present method works, we now consider
a few specific cases.
\medskip
\noindent{\bf 5.1. Coulomb Fluid in a Half Space}
\medskip
The surface of the Coulomb fluid is the plane $x 0y$, which acts as
an impenetrable wall, confining the particles to the half-space $z <
0$. The method of images gives, in the empty half-space,
$$
G({\bf r }, {\bf r'}) = {1 \over |{\bf r} - {\bf r'}|} - {1\over
|{\bf r}^\star - {\bf r'}|} \eqno (5.1)
$$
where ${\bf r}^\star = (x, y, -z)$ is the image of ${\bf r} = (x, y,
z)$. One readily finds from (4.3), on the plane $x 0y$,
$$
\beta \langle  \sigma({\bf r}) \sigma ({\bf r'}) \rangle^T = - {1
\over 8 \pi^2 |{\bf r} - {\bf r'}|^3} \eqno (5.2)
$$
which is the result which had been obtained in ref. 9 in the
equivalent microscopic language~: the volume charge density
correlation function $\langle \rho ({\bf r}) \rho ({\bf r'})
\rangle^T$ behaves asymptotically when $(x - x')^2 + (y -y')^2$
becomes large as $F(z, z', (x-x')^2+(y-y')^2)$, where $F$ is a
function integrable upon $z$ and $z'$ obeying
$$
\beta\int^0_{-\infty} dz \int_{-\infty}^0 dz' F(z, z',
(x-x')^2+(y-y')^2) = - {1\over 8 \pi^2[(x-x')^2+(y-y')^2]^{3/2}} \eqno
(5.3) $$
\medskip
\noindent{\bf 5.2. Coulomb Fluid in a Ball}
\medskip
A ball of macroscopic radius $R$, centered at the origin, is filled
by a Coulomb fluid. The method of images now gives, outside the ball,
$$
G({\bf r}, {\bf r'}) = {1 \over |{\bf r}- {\bf r'}|} - {R \over r}
{1 \over |{\bf r'}- {\bf r}^\star|} + {R \over rr'} \eqno (5.4)
$$
where the image of ${\bf r}$ has the coordinate ${\bf r}^\star =
R^2{\bf r}/r^2$. The presence of the last term in (5.4) is
appropriate for an insulated ball, i.e. if we want to study the
fluctuations in an ensemble such that the total charge on the ball
is constant, for instance the canonical ensemble; on the contrary, if
the ball is assumed to be allowed to exchange charge with some
reservoir, say at zero potential (grounded ball), the last term of
(5.4) must be omitted.

Using (5.4) in (4.3) gives, for an insulated ball,
$$
\beta \langle \sigma ({\bf r}) \sigma ({\bf r'}) \rangle^T = - {1
\over 8 \pi^2} [ {1 \over (2R \sin {\theta\over 2})^3} + {1 \over
2R^3}] \eqno (5.5)
$$
where $\theta$ is the angle between ${\bf r}$ and ${\bf r'}$. In the
case of a grounded ball, the last term in (5.5) must be suppressed,
i.e. one must add the contribution $+1/(4\pi)^2 R^3$ from the total
charge fluctuations.

This contribution from the fluctuations of the total charge $Q$ can
be alternatively derived by assuming that, for typical
configurations, $Q$  is uniformly spread on the surface. Then, the
corresponding energy is $Q^2/2R$. If one further assumes that the
average of this energy has the harmonic oscillator value
$(1/2)\beta^{-1}$, using $\sigma = Q/4\pi R^2$ one finds the
contribution $1/(4\pi)^2R^3$ to $\beta \langle\sigma \sigma \rangle$.
\medskip \noindent{\bf 5.3. Coulomb Fluid in a Wedge}
\medskip
Our method reproduces the results of Choquard {\it  et al.}$^{(8)}$.
These authors had expressed some doubts about the
reliability of their result, because it was not obvious that their
derivation, presented as an approximation to the Debye-H\"uckel
theory, was valid for a non-smooth surface such as a wedge. The
present approach now puts things on the firmer basis of macroscopic
electrostatics, that we believe to be true, even in the case of a
wedge.

The results in ref. 8 are different from the ones which had been
obtained by Jancovici {\it et al}.$^{(10)}$ It is now apparent that
the Section Wedge in ref. 10 is erroneous\footnote{$^3$}{The mistake
in ref.10 can be traced to the assumption following eq.(A.5); this
assumption is just not true.}, as well as its quotation in eqs. (10a)
and (10b) of ref. 3. \bigskip
\noindent {\bf 6. CONDUCTING SURFACES}
\bigskip
This Section is about two-dimensional Coulomb fluids, with an
interaction potential which is the usual $1/r$ one. The surface
charge correlations can be studied by a slightly modified version of
the method which was used in Section 4. The field correlation
function in three-dimensional space is still given by (3.3), and the
surface charge correlations can be still expressed in terms of the
correlations between the field discontinuities on the surface.

In the case of a conducting plane, one finds
$$
\beta \langle \sigma ({\bf r}) \sigma ({\bf r}')\rangle^T = - {1\over 4
\pi^2 |{\bf r}-{\bf r}'|^3} \eqno (6.1)
$$
(This result could also have been derived by considering a
slab$^{(3,10)}$ in the zero thickness limit). It should be noted that
the plane result (6.1) is not identical to the half-space result
(5.2). This shows that, in the case of a three-dimensional Coulomb
fluid with boundaries, there must be some coupling between the
surface charge and the volume charge densities.

In the case of an insulated conducting spherical surface of radius
$R$ centered at the origin, one finds
$$
\beta \langle \sigma ({\bf r}) \sigma ({\bf r}')\rangle^T = - {1\over
4\pi^2} [ {1 \over (2R \sin {\theta\over 2})^3} + {1\over 4R^3}] \eqno
(6.2)
$$
where $\theta$ is the angle between ${\bf r}$ and ${\bf r'}$.
In the case of a grounded surface, the last term of   (6.2) must be
suppressed. Therefore, $1/(4\pi)^2R^3$ is again the contribution from
the total charge fluctuations, just as for a grounded ball filled with
a three-dimensional Coulomb fluid, in agreement with the assumption
that in this latter case the fluctuations of the total charge are
localized on the surface only. The rest of $\langle \sigma ({\bf r})
\sigma ({\bf r}')\rangle^T$ however is different for a spherical surface
and for a ball filled with the fluid, as seen in (6.2) and (5.5);
again, in the case of a ball, there is some coupling between the
surface and the volume.
 \bigskip \noindent{\bf 7. TWO-DIMENSIONAL
COULOMB INTERACTION} \bigskip
The above method and results can be easily transposed to
two-dimensional or one-dimensional systems with a two-dimensional
Coulomb interaction : the potential at ${\bf r}$ created by a unit
point charge at the origin is $-\ln(r/L)$, where $L$ is some given
length. These systems are interesting for a variety of reasons. In
some cases they are exactly solvable models. Some of them appear in
the theory of random matrices.

One now obtains for the potential-potential correlations inside an
insulated two-dimensional conductor
$$
\beta \langle \phi ({\bf r}) \phi({\bf r}') \rangle^T = - \ln {|{\bf r}
- {\bf r}'| \over L} - {1\over C}
\eqno (7.1.)
$$
where $C$ is the capacitance; for instance, for a disk of radius
$R$, $C^{-1} = - \ln (R/L)$. Now, if the system becomes infinite,
$C^{-1} \rightarrow \infty$ and (7.1) has no well-behaved
thermodynamic limit.$^{(11)}$ However, the corresponding field-field
correlation function is given by
$$
\beta \langle E_\mu({\bf r}) E_\nu ({\bf r}') \rangle^T = - {2 ({\bf r} -
{\bf r}')_\mu ({\bf r} - {\bf r}')_\nu - \delta_{\mu\nu}({\bf r} - {\bf
r}')^2\over
|{\bf r} - {\bf r}'|^4} \eqno (7.2)
$$
while $\langle E_\mu({\bf r})\rangle = 0$, and these expressions remain
finite for infinite systems.
The analog of the surface charge density is now a charge per unit
length that we still call $\sigma({\bf r})$. Following the same other
steps as in Sections 2, 3, 4, one finds again that obtaining the
``surface'' charge correlations $<\sigma \sigma >$ for two-dimensional
or one-dimensional systems with logarithmic interactions
reduces  (for macroscopic distances) to a macroscopic electrostatics
problem : finding $G({\bf r}, {\bf r}')$, the total potential at ${\bf
r}'$ outside a conductor when a unit point charge has been put at
${\bf r}$ (also outside). For the explicit calculation of $G$, a
specifically two-dimensional tool is provided by the theory of
functions of a complex variable and conformal transformations.

For a Coulomb fluid in a half plane (the fluid is assumed to
fill the $y < 0$ domain of the $xy0$ plane), one finds
$$
\beta \langle \sigma(x) \sigma (x') \rangle^T = - {1 \over
2\pi^2(x-x')^2} \eqno (7.3)
$$
which is the result which had been given in ref. 9 in microscopic
language. Similarly, on an infinite conducting line, one finds
$$
\beta \langle \sigma (x) \sigma (x') \rangle = - {1\over
\pi^2(x-x')^2} \eqno (7.4)
$$
(a result which had been previously derived by a different
method$^{(12)}$).

For a Coulomb fluid in a disk of radius $R$ centered at the origin,
one finds
$$
\beta \langle \sigma({\bf r}) \sigma ({\bf r}') \rangle = - {1 \over
2\pi^2(2R \sin {\theta\over 2})^2} \eqno (7.5)
$$
where $\theta$ is the angle between ${\bf r}$ and ${\bf r'}$. This
result is valid both for an insulated and a grounded disk. In the
case of a three-dimensional ball, grounding gave
to the surface charge correlation function (5.5) an additional
contribution coming from the fluctuations of the total charge. It
is remarkable that such an additional term does not occur in two
dimensions. The reason is that the energy associated to a total
charge fluctuation $Q$ on a disk is infinite, since increasing
the charge $Q$ by bringing in an additional charge $\delta Q$
from far away costs an energy $Q\delta Q \int^\infty_R dr/r$,
where the integral diverges. Thus, fluctuations of the total charge
cannot occur.

As an illustration of the method of conformal transformations, let us
consider a Coulomb fluid in a wedge. We use polar coordinates $(r,
\theta)$. The fluid is assumed to fill the two-dimensional wedge
$\gamma < \theta < 2\pi$; the domain $0 < \theta < \gamma$ is empty.
For finding $G$ in the wedge $0 < \theta < \gamma$, we can start
from its half-plane $(\gamma = \pi)$ expression
in terms of complex coordinates $[z = r \exp(i\theta)]$
$$
G_{\gamma = \pi} = \ln |{\bar z -z'\over z - z'}| \eqno (7.6)
$$
and make the conformal transformation $z \rightarrow
z^{\pi/\gamma}$ which relates the half plane and the wedge. Thus,
$$
G_\gamma = {1\over 2} \ln {r^{2\alpha} + r'^{2\alpha} - 2
r^\alpha r'^\alpha \cos \alpha(\theta + \theta')\over r^{2\alpha} +
r'^{2\alpha} - 2
r^\alpha r'^\alpha \cos \alpha(\theta - \theta') } \eqno (7.7)
$$
where $\alpha = \pi/\gamma$, and one finds
$$
\beta \langle \sigma({\bf r}) \sigma ({\bf r}') \rangle = -
{\alpha^2\over 2\pi^2} {(rr')^{\alpha-1}\over (r^\alpha -
sr'^\alpha)^2} \eqno (7.8)
$$
where $s = +1(-1)$ if the two points ${\bf r}$ and ${\bf r}'$ are on
the same side (different sides) of the wedge.

Other specific cases are solved in refs. 13 and 14.
\bigskip
\noindent{\bf 8. GAUSSIAN BEHAVIOR OF THE POTENTIAL AND FIELD
FLUCTUATIONS}
\bigskip
In Sections 2 and 3, we have obtained the two-point correlations
for the electrical potential and field, except for their
microscopic distance behavior. We now want to prove the stronger
statement that the fluctuations of these quantities are
``almost'' Gaussian, with the above mentioned covariances (the
meaning of ``almost'' will be explained hereafter).

Showing that the fluctuations are Gaussian is a step towards better
understanding why the finite-size corrections to the free energy of a
Coulomb system$^{(15)}$ are similar to those of a critical field theory
: the Gaussian model.

The basic ingredient of the proof (already used in wave-number
space in ref.15) is the assumption that the
screening response (2.3) remains of the same form, {\it linear} in $q$,
even if $q$ is no longer infinitesimal but is an external point
charge of arbitrary magnitude.  Now, a
{\it linear} response to a {\it finite} (non
infinitesimal) perturbation characterizes a Gaussian distribution. As a
pedagogical example, consider a particle in a one-dimensional
potential $V(x)$. A simple calculation shows that  if, in thermal
equilibrium, the average displacement of the particle under an
additional force $q$ is proportional to $q$, then $V(x)$ is a harmonic
oscillator potential, and the probability density $\exp[- \beta V(x)
]$ is Gaussian.

For a proof  in the present case of a
Coulomb fluid, we study the cumulants, considering for instance an
insulated fluid with three-dimensional Coulomb interactions. The
argument is as follows. Place at $n-1$ points ${\bf r}_1, {\bf r}_2,
..., {\bf r}_{n-1}$ (chosen inside or outside the fluid) point charges
$q_1, q_2, ..., q_{n-1}$. Let ${\bf r}_n$ be some point, inside or
outside; if inside, ${\bf r}_n$ is assumed to be at a distance large
compared to the microscopic scale from those points among $({\bf r}_1,
{\bf r}_2, ..., {\bf r}_{n-1})$ which are inside. Since macroscopic
electrostatics says that the potential-charge relations are linear,
the response of $\phi({\bf r}_n)$ is given by an obvious
generalization of what has been derived in Sections 2 and 3 :
$$
\langle \phi ({\bf r}_n) \rangle_{\{q_i\}} = \langle
\phi({\bf r}_n)\rangle + \sum^{n-1}_{i=1} a_i q_i \eqno (8.1)
$$
where $\langle\ \rangle_{\{q_i\}}$ means an average computed with
the full Hamiltonian including the interaction term
$\sum^{n-1}_{i=1} q_i \phi({\bf r}_i)$, while $\langle\ \rangle$ means
the average when there are no charges $ q_i$. The coefficients $a_i$
are those ones obtained in Sections 2 and 3 :
$$
\eqalignno{
a_i & = -{1\over |{\bf r}_i - {\bf r}_n|} + {1\over C}\quad \hbox{if both
${\bf r}_i$ and ${\bf r}_n$ are inside}\cr
a_i & = -{1\over |{\bf r}_i - {\bf r}_n|} + F({\bf r}_{\hbox{out}}) \quad
\hbox{if one member $({\bf r}_{\hbox{out}})$ of the pair $({\bf r}_i,
{\bf r}_n)$}\cr
&\hbox{ is outside and the other one inside}\cr
a_i & = -{1\over |{\bf r}_i - {\bf r}_n|} + G({\bf r}_i, {\bf r}_n)
\quad\hbox{if both ${\bf r}_i$ and ${\bf r}_n$ are outside}&(8.2)\cr}
$$
By definition,
$$
\langle \phi({\bf r}_n)\rangle_{\{q_i\}} = {\langle e^{-\beta
\sum_{i=1}^{n=1} q_i \phi({\bf r}_i)}\phi({\bf r}_n)\rangle\over \langle
e^{-\beta \sum^{n-1}_{i=1} q_i \phi({\bf r}_i)}\rangle} \eqno (8.3)
$$
Rewriting the r.h.s. of (8.3) as a logarithmic derivative  and
combining (8.1) and (8.3), one obtains
$$
- {1\over \beta} {\partial \over \partial q_n} \ln \langle
e^{-\beta \sum^n_{i=1} q_i \phi ({\bf r}_i)} \rangle \big |_{q_n = 0} =
\langle \phi({\bf r}_n)\rangle + \sum^{n-1}_{i=1} a_i q_i \eqno (8.4)
$$
Since the cumulants (or truncated correlation functions) $\langle
\ldots \rangle^T$ of the potential $\phi$ are defined by the
expansion in powers of $q_i$
$$
\ln \langle e^{-\beta \sum^n_{i=1} q_i \phi ({\bf r}_i)}\rangle =
\sum_0^\infty{}' {\langle \phi({\bf r}_1)^{m_1} \phi({\bf r}_2)^{m_2} \ldots
\phi({\bf r}_n)^{m_n} \rangle^T\over m_1 ! m_2 ! \ldots m_n !}
(-\beta q_1)^{m_1} (-\beta q_2)^{m_2} \ldots (-\beta q_n)^{m_n}
\eqno (8.5)
$$
(where the prime indicates the absence of the term with all $m$'s
simultaneously vanishing), a comparison between (8.4) and (8.5)
shows that all cumulants of total order $m_1 + m_2 + \ldots + m_n$
higher than $2$ and with $m_n = 1$ vanish.

The above result can be rephrased as follows. Consider the
cumulant $\langle\phi({\bf r}_1) \phi({\bf r}_2)$
$\ldots\phi({\bf r}_p)\rangle^T$, with $p > 2$. That cumulant
vanishes provided that one point at least, say ${\bf r}_p$, is at a
distance large compared to the microscopic scale from all those
other points which are inside the fluid; these other points
however might be at arbitrary distances (including zero) from each
other. Similar properties hold for the electrical fields.

Were it not for the restriction about the position of ${\bf r}_p$,
the vanishing of the cumulants of order larger than 2 would imply
that the $\phi({\bf r}_i)$ are jointly Gaussian. With the restriction
added, the statistics become ``generalized Gaussian''. This
restriction actually is not very drastic, or at least is of a kind
which currently occurs in field theory. In the simple case of a
homogeneous infinite system, the covariance $\beta \langle
\phi(r_1) \phi(r_2) \rangle = 1/|{\bf r}_1 - {\bf r}_2|$ corresponds
to a Gaussian probability density proportional to $\exp [-(\beta/8\pi)
\int ({\bf \nabla} \phi)^2 d^3 {{\bf r}}]$, a form which is not
unexpected since $(1/8\pi) ({\bf \nabla}\phi)^2$ is just the
Coulomb energy density expressed in terms of the electrical field
$- {\bf \nabla} \phi$. It is well-known that this Gaussian field
theory actually becomes singular for short distances, and the
average $\langle \phi({\bf r}_1) \phi({\bf r}_2) \ldots \phi
({\bf r}_p)\rangle$ is not properly defined when there are coincident
points; some regularization is necessary.  A most important property
is that, at least in the absence of coincident points (i.e. when all
distances are large compared to the microscopic cutoff), Wick's
theorem can be straightforwardly used for expressing $\langle
\phi({\bf r}_1) \phi ({\bf r}_2) \ldots \phi ({\bf r}_p) \rangle$ in
terms of the 2-point function $ \langle \phi({\bf r}_1) \phi({\bf
r}_2) \rangle = 1 / \beta |{\bf r}_1 - {\bf r}_2|$.

The Gaussian behavior of the electrical fields follows.

Our result about the higher-order cumulants had been previously
obtained, by a different method, in the special case of an
infinite homogeneous one-component plasma.$^{(16)}$ It should be
noted that the quantities $V^{\hbox{out}}_R$ which  were shown to
have Gaussian fluctuations in ref. 6 are {\it not} the full
potentials $\phi$ considered here.

The present considerations can be easily transcribed to the case
of a two-dimensional (logarithmic) Coulomb potential.
\bigskip
\noindent {\bf 9. CONCLUSION}
\bigskip
The universal validity of macroscopic electrostatics implies in
particular universal response properties for conductors. Whenever
{\it classical} statistical mechanics is applicable, these
response properties imply in turn universal properties for the
equilibrium thermal fluctuations. Independently of the detail of
the microscopic constitution of the conductor, for length scales
large compared to the microscopic ones, the potential and field
fluctuations are Gaussian with universal covariances, (outside the
conductor, things depend on its geometry, but again not on its
microscopic constitution). The surface charge fluctuations are
also universal, depending only on the geometry. The volume charge
fluctuation also have universal properties, but in a less explicit
way, since the universality then appears only in the form of sum
rules.

Unfortunately, when quantum mechanics
is used,the universal response properties no longer imply simple
universal static (equal time) correlations, because of the more
complicated form of the linear response theory. One can write quantum
sum rules for the correlations in the special case of the one-component
plasma$^{(17, 10)}$, but these sum rules involve the non-universal
plasma frequency. For more general quantum models, no sum rules are
known for static or (real) time-dependent correlations.
\vfill\eject

\noindent {\bf REFERENCES}
\bigskip
\item{1.}  Ph. A. Martin, Rev. Mod. Phys. {\bf 60} : 1075 (1988).

\item{2.} F.H. Stillinger and R. Lovett, J. Chem. Phys. {\bf 49}
: 1991(1968).

\item{3.} B. Jancovici, in {\it Strongly Coupled Plasma Physics}, F.J.
Rogers and H.E. DeWitt, eds. (Plenum, New York, 1987).

\item {4.} J.M. Kosterlitz and D.J. Thouless, J. Phys. C{\bf 6} :
1181 (1973).

\item{5.} H. Schultz, J. Phys. A{\bf 14} : 3277 (1981).

\item{6.} J.L. Lebowitz and Ph. A. Martin, J. Stat. Phys. {\bf
34} : 287 (1984).

\item{7.} S.L. Carnie and D.Y.C. Chan, Chem. Phys. Lett. {\bf 77}
: 437 (1981).

\item{8.} Ph. Choquard, B. Piller, R. Rentsch, and P.
Vieillefosse, J. Stat. Phys. {\bf 55} : 1185 (1989).

\item{9.} B. Jancovici, J. Stat. Phys. {\bf 29} : 263 (1982).

\item{10.} B. Jancovici, J.L. Lebowitz, and Ph. A. Martin, J.
Stat. Phys. {\bf 41} : 941 (1985).

\item{11.} A. Alastuey and B. Jancovici, J. Stat.Phys. {\bf 34} :
557 (1984).

\item{12.} P.J. Forrester, B. Jancovici, and E.R. Smith, J. Stat.
Phys. {\bf 31} : 129 (1983).

\item{13.} B. Jancovici and P.J. Forrester, Phys. Rev. B, to be
published.

\item{14.} P.J. Forrester, Nuclear Phys. B, to be published.

\item{15.} B. Jancovici, G. Manificat, and C. Pisani, J. Stat.
Phys. {\bf 76} : 303 (1994).

\item{16.} L.G. Suttorp, J. Stat. Phys. {\bf 66} : 1343 (1992).

\item{17.} D. Pines and Ph. Nozi\`eres, {\it The Theory of Quantum
Liquids} (Benjamin, New York, 1966).
 \bye